\documentclass[preprint,11pt]{revtex4-1}

\usepackage{siunitx}
\usepackage{amsmath}
\usepackage{amsfonts}
\usepackage{graphicx}
\usepackage{xcolor}
\usepackage{siunitx}

\usepackage{ulem}
\usepackage{comment}

\newcommand{\Dunit}{\si{\micro\metre\squared\per\second}}

\newcommand{\nm}{\si{\nano\metre}}

\begin{document}

\title{Thermofluidic non-equilibrium assembly of reconfigurable functional structures}%

\author{Desmond Quinn, Diptabrata Paul, Frank Cichos}%
\email[Frank Cichos ]{cichos@physik.uni-leipzig.de}
\affiliation{Molecular Nanophotonics Group, Peter Debye Institute for Soft Matter Physics, Leipzig University, 04103 Leipzig, Germany}

\begin{abstract}
Non-equilibrium assembly, driven by fluxes controllable by continuous external energy inputs, enables dynamic and reconfigurable structures. Such controlled 3D assembly is desired for the design of adaptive materials that exploit structure-function relationships, but has remained challenging. We present a non-equilibrium assembly of colloidal particles mediated by laser-induced local heating and continuous heat dissipation. These 3D out-of-equilibrium structures, assembled in a matter of a few minutes, were highly ordered and exhibited tunable photonic stopbands. We quantify the particle fluxes from the underlying assembly processes and report the growth dynamics of the assembled structures. Furthermore, we demonstrate the modulation of the photonic stopband achieved by modulating the particle fluxes, highlighting the prospects of such thermofluidic assembly for creating reconfigurable functional structures.
\end{abstract}
\maketitle

\section*{Introduction}
The assembly of simple components into functional structures is a fundamental process across different scales. In biological systems, this is demonstrated through the remarkable self-organization of cellular components into complex living structures \cite{alberts2017molecular, hess2017non}. In the realm of synthetic materials, simpler yet functional structures can be created through the assembly of engineered units, such as dielectric components that may form, for example, metamaterials \cite{huh2020exploiting}. These synthetic assembly processes are typically driven toward equilibrium states through free energy minimization \cite{rocha2020role, sciortino2019entropy}. Colloidal particle assemblies exemplify this equilibrium-driven approach \cite{li2021colloidal}, offering various functional capabilities \cite{boles2016self} that enable applications in photonics \cite{kim2011self}, electronics \cite{zhu2017assembly}, chemistry \cite{wintzheimer2018supraparticles}, and lithography \cite{zhang2009colloidal}.
Alternatively, assembly can be also driven by continuous energy dissipation in non-equilibrium processes, leading to stationary states that are maintained by the energy flux. According to Prigogine's theorem \cite{prigogine2017non}, such systems tend toward states of minimum entropy production while maintaining their non-equilibrium character, allowing the dynamic creation and destruction of structures not feasible with equilibrium assembly \cite{heinen2015celebrating, arango2019self}. These dissipative structures exhibit a well-defined response to external perturbations, leading to new functionalities. For example, inhomogeneous temperature fields created by laser heating can drive the assembly of colloidal structures that exhibit random lasing, where the lasing properties can be reconfigured by modulating the laser power \cite{Trivedi.2022}. While techniques enabling local control over assembly through energy flux have remained challenging, they are particularly valuable as they offer the potential to create adaptive structures similar to those found in biological systems. Such approaches could enable reconfigurable structures across diverse scales - ranging from nanoparticles and biological molecules \cite{hess2017non} to living cells and bacteria.
Chemical approaches to non-equilibrium assembly, while powerful in their use of dissipative reaction networks \cite{van2020out, heinen2015celebrating, van2017dissipative}, are often limited to specific systems and face challenges in achieving precise spatial control. This has motivated the exploration of physical control methods. Particularly, assembly processes driven by external electric and magnetic fields \cite{grzelczak2010directed, bharti2015assembly, ma2013formation, al2020magnetic} have emerged as an alternative approach, though they too face challenges, primarily in achieving local rather than bulk control over the assembly process.

Temperature gradients created by localized optical absorption drive persistent flows through continuous energy dissipation \cite{bregulla2016thermo, franzl2022hydrodynamic, jiang2009manipulation, maeda2011thermal, liu2021opto}, offering a versatile approach to non-equilibrium assembly. This optical control enables highly localized manipulation of matter and, unlike chemically-driven systems, can be applied to a broad range of materials. The interplay between different thermally-driven processes - such as thermophoresis, thermoosmosis, and convection - creates complex dynamics that can lead to diverse self-organized structures. Understanding the interplay of these effects, is therefore a key to controlling the assembly process and the resulting structures.

Here we report experiments to understand the interplay of thermoosmosis, thermophoresis, and temperature induced depletion for the dynamic formation of colloidal structures out-of-equilibrium. We observe colloidal crystal formation under the osmotic pressure generated by the thermophoresis of polyethylene glycol (PEG) molecules and the hydrodynamic forces generated by the thermoosmotic flow. The crystal formation happens despite the thermophoretic repulsion of the colloids from the heat source and has been verified by the formation of a photonic stop band by angle resolved spectroscopy. At higher temperatures, the thermoosmotic flows break the radial symmetry of the forces and lead to the formation of a dynamic toroidal structure that can be collapsed into the crystal structure by a rapid quenching process.
Our approach achieves a high degree of control over the assembled structures through time-variable temperature fields induced by lasers in highly localized regions. We explore this structure formation process and report on the emergence of functional photonic properties in the resulting assemblies. These findings provide new insights into controlling non-equilibrium assembly processes and could enable the development of reconfigurable photonic devices.
%
%
\section*{Results and Discussion}

\begin{figure*}[!htb]
    \centering
    \includegraphics[width=0.9 \textwidth]{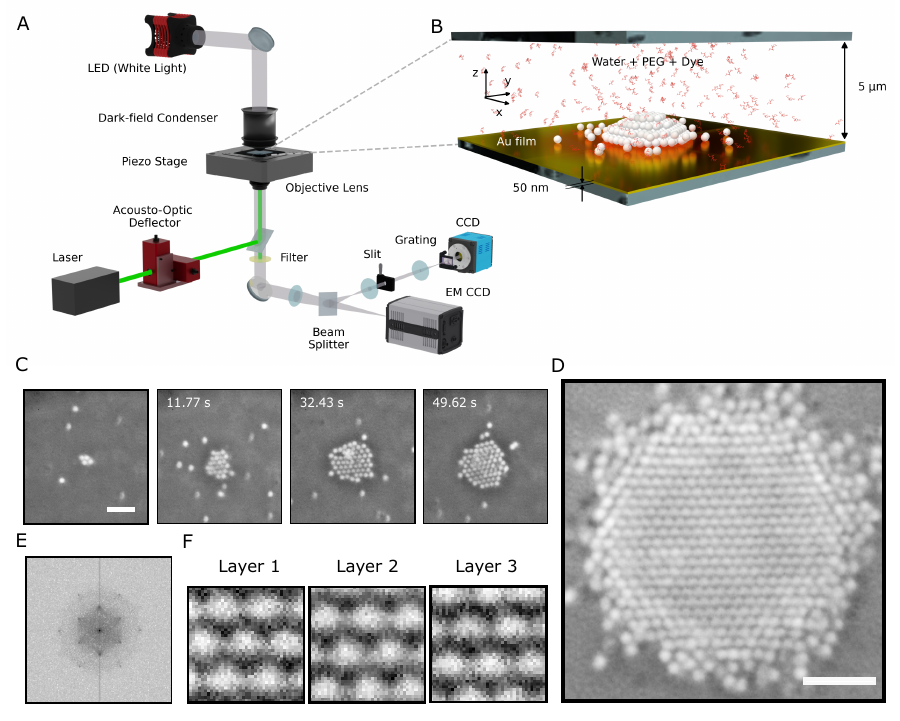}
    \caption{\textbf{Thermofluidic assembly of a 3D colloidal crystal}. Polystyrene (PS) particles $442 \, \si{\nano\metre}$ are assembled by laser induced heating (A) The experimental setup consists of a steerable $532 \, \si{\nano\metre}$ laser in
    an inverted microscope that images the sample in both Real and Fourier space, with the Fourier imaging path containing a flippable slit and grating for measuring spectra. (B) The sample consists of a $5 \, \si{\micro\metre} $ thick liquid film (water + PEG) between two glass coverslips. The lower glass slides is coated with a Gold film ($50 \, \si{\nano\metre}$), which is heated by a focused laser. (C) A timeline of the assembly of the colloidal crystal. (D) An image of the highly ordered crystal. (E) Fourier transform of the preceding microscopy image displaying a sixfold symmetry (F) The different layers of the colloidal crystal, imaged by focusing through the structure with the piezo stage }\label{fig:figure1}
\end{figure*}

\textbf{Thermofluidic Assembly} The assembly was carried out in an inverted microscope setup that uses an Acousto-Optic Deflector (AOD) to control and modulate a $532\, \si{\nano\metre}$ wavelength laser spot that was focused to the sample plane by a oil-immersion objective lens (100x, NA 0.5-1.35). The focused laser spot enables local heating of a gold film ($50 \, \si{\nano\metre}$) by optical absorption. The resulting temperature was calibrated using a liquid crystal method \cite{franzl2022hydrodynamic} and substantiated using finite-element simulations (detailed in the SI). The sample was imaged in real space using dark-field illumination and in Fourier space via a Bertrand lens \cite{kurvits2015comparative} in two separate imaging paths. The Fourier space image which corresponds to the Back Focal Plane (BFP) of the objective lens contains the angle resolved information for a wavelength range selected by a variable band pass filter placed along the path. Further, an adjustable slit and flippable grating added to the Fourier imaging path was used to obtain spectroscopic information by dispersing the light before being captured by the camera.  (see Fig.~\ref{fig:figure1}A).
The assembly was carried out for Polystyrene (PS) particles ($345\, \si{\nano\metre}$, $400\, \si{\nano\metre}$, and $442\, \si{\nano\metre}$ diameter; measured via Dynamic Light Scattering (DLS)) in a solution of 7 $\%$ (w/v) Polyethylene glycol 6000 (PEG). The colliodal solution was placed between two glass coverslips maintained at $5\,\si{\micro\metre}$ distance to  facilitate growth in 3 dimensions. (see Fig.~\ref{fig:figure1}B)

\noindent
\textbf{Experimental growth dynamics of the colloidal assemblies}
In the absence of any heating of the gold film, the $442\, \si{\nano\metre}$ PS particles carry out Brownian motion with a diffusion coeffcient of $D=0.97\, \Dunit$. Increasing the laser power to $P_0=0.26\, \si{\milli\watt}$ leads to a local heating of the gold film and a steady radial influx of particles yielding the assembly of a crystalline 2-dimensional(2-d) colloidal structure at the gold surface as displayed examplarily in Fig.~\ref{fig:figure1}C for $420\, \nm$ PS particles. When a larger 2d structure has formed, particles also start to occupy the interstitial sites on top of the 2-d layer to form a 3-dimensional (3-d) crystal. A highly crystalline order was observed (see Fig.~\ref{fig:figure1}D and E). The crystal structure is well identified in the optical images for $442 \,\si{\nano\metre}$ particles and reveals an FCC stacking as measured by focusing through the crystal structure using a piezo-electric stage(see Fig.~\ref{fig:figure1}F). The measured distance between successive layers in the assembled structures also corresponds to $d_{111}$, the spacing between the (111) lattice planes of such an FCC crystal. When the heating laser is turned off, the crystal structure disintegrated, causing the particles to disperse through Brownian motion.

\begin{figure*}[!htb]
    \centering
    \includegraphics[width=0.9 \textwidth]{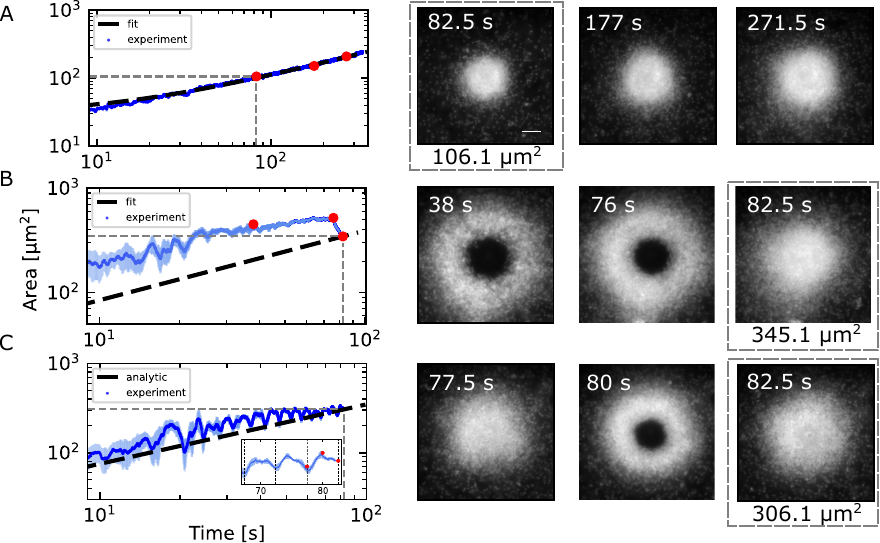}
    \caption{\textbf{Growth dynamics for different heating modes}. The growth of the area of the structure directly above the gold film was monitored over time. $345 \, \si{\nano\metre}$ PS particles were used. (A) Colloidal structure was assembled with a constant laser power $P_0 = 0.26\,\si{\milli\watt}$. (B) Colloidal structure was assembled at a higher laser power of $P_1 = 2.16\,\si{\milli\watt}$ (for $76 \, \si{\second}$, before being collapsed back into a crystal structure by lowering the laser power back to $P_0$. (C) Colloidal structure was assembled with a pulsed laser intensity, cycling between the higher laser power $P_1$ and lower laser power $P_0$ with a period of $10\, \si{seconds}$. The sample images show the switching of the structures at points of low and high laser power. The inset shows the area of the structure oscillating with well defined periods at larger times. The period corresponded to that of the laser pulse. The plotted experimental data considers a moving average with window size of 50 frames.  }\label{fig:figure2}
\end{figure*}

The dynamics of the crystal growth depends on the heating power and can be characterized by measuring the change of the area occupied by the structure as a function of time.
Fig.~\ref{fig:figure2}A illustrates the growth dynamics of the assembled structure for $345\,\si{\nano\metre}$ PS particles at an incident laser power of $P_\text{0} = 0.26\,\si{\milli\watt}$, which corresponds to a maximum temperature increment of $\Delta T_\text{max} = 6\,\si{\kelvin}$ directly at the gold film.
The area $A$ was found to grow as a power law given by:
\begin{equation}\label{eq1}
 A(t)=[k(t+t_0)]^{\alpha}
\end{equation}

having exponent $\alpha = \frac{2}{3}$ (Fig. \ref{fig:figure2}A). A growth rate $k=11.16\,\si{\micro\metre\cubed\per\second}$ was obtained for the corresponding growth by fitting the measured area to the aforementioned power law. After $82.5 \, \si{\second}$, a crystal structure with an area of $A=106.1 \, \si{\micro\metre\squared}$ and several layers was formed.

A further rise in laser power and temperature results in a stronger radial influx of particles and accelerated growth. Above a certain temperature increment of $\Delta T_\text{c}=10\,\si{\kelvin}$, however, there are qualitative differences, as the resulting ensemble structure is no longer a solid crystal, but highly dynamic (see video in the additional information). A circular zone opens up in the center where the particles get depleted. The radius of which further increases with increasing heating power.

Fig.~\ref{fig:figure2}B displays the growth of an assembled structure of $345\, \si{\nano\metre}$ PS particles at a constant laser power of $P_\text{1} = 2.1\, \si{\milli\watt}$ ($\Delta T_\text{max} = 48.8\, \si{\kelvin}$). The image at $t=38 \, \si{\second}$ depicts the hole of $5.6\, \si{\micro\metre}$ radius at the center of the assembly, while at $t=76 \, \si{\second}$, the assembly appears significantly denser with a smaller hole ($4.4\, \si{\micro\metre}$ radius) and a larger surface area.
Upon decreasing the laser power back to $P_\text{0} = 0.26\,\si{\milli\watt}$, the structure collapses into a dense ensemble after a few seconds (see Fig.~\ref{fig:figure2}B) in a rapid quenching process. The final area after $t=82.5\, \si{\second}$ was $345.1\, \si{\micro\metre\squared}$, 3.25 times higher than the area at the same time point assembled with power P0. This final area of the structure corresponds to an effective growth rate $k=77.7\,\si{\micro\metre\cubed\per\second}$ , extracted from the power law proposed earlier.

The rapid quenching was repeated in a cyclic manner to switch between a highly dynamic toroidal structure and a dense solid structure by switching between the heating powers $P_\text{0}$ and $P_\text{1}$ as indicated in Fig. \ref{fig:figure2}C for a cycle time of $5\,\si{\second}$. During cycles of low and high heating power, the structure is still growing.

After 82.5 seconds, the collapsed state covers an area of $306.1\, \si{\micro\metre\squared}$, which is 2.9 times higher than the steady growth case, and corresponds to an effective growth rate of $k=64.91\,\si{\micro\metre\cubed\per\second}$. At larger times, the corresponding growth curve contains periodic oscillations with frequency corresponding to the cycle time of the laser power (see inset of Fig.\ref{fig:figure2}C). The appearance of prominent peaks suggest that once a large number of particles have accumulated, a dense, 3-dimensional structure can be formed and destroyed at the center in a matter of seconds. 
\\

\noindent
\textbf{Emergence of photonic property} Distinct properties can emerge when colloidal nanoparticles are assembled into ordered structures. Dielectric structures exhibiting spatial periodicity modulate the propagation of electromagnetic waves due to the photonic band structure dictated by their crystal structure. In particular, stop bands, i.e. directions in which light propagation is forbidden for certain wavelengths, can emerge.
\begin{figure*}[!htb]
    \centering
    \includegraphics[width=0.9 \textwidth]{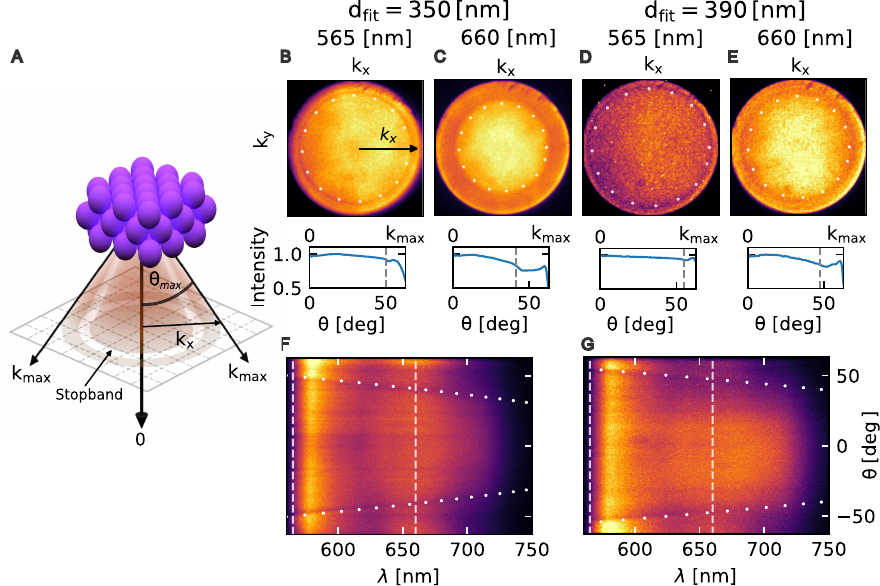}
    \caption{\textbf{Photonic stopband in the assembled structures}. (A) Illustration of light from embedded emitters propagating through the assembled crystal. (B and C) BFP images (and corresponding radial profile) of the assembled $345\, \si{\nano\metre}$ PS particles for the specified optical wavelengths. (D and E) BFP images (and corresponding radial profile) of the assembled $400\, \si{\nano\metre}$ PS particles for the specified optical wavelengths. (F) BFP spectra of the assembled $345\, \si{\nano\metre}$ PS particles. (G) BFP spectra of the assembled $400\, \si{\nano\metre}$ PS particles. The fitted diameters reported in the figure match closely with the theoretical prediction} \label{fig:figure3}
\end{figure*}
To experimentally determine the stopbands in our assembled structures, we accessed the back focal plane (BFP) of the objective lens which provides an angle-resolved measure of the collected light \cite{wagner2013fast}. Dye molecules embedded inside the photonic crystal with a suitable absorption/emission characteristics provided the necessary emission inside the crystal (see Methods). The excited fluorescent light of these internal emitters propagates through the assembled crystal structure and is modified by its photonic properties (Fig.~\ref{fig:figure4}A).

The examined BFP images of the dye embedded assembled crystals contained evidence of altered photonic properties. The BFP images (Fig.~\ref{fig:figure4}D-E), exhibited dark circular bands corresponding to certain wavevectors (angles), implying the presence of photonic stopbands in the assembled structures. The radial intensity profiles further highlight this dip in intensity.
Using different optical filters shifted the position of the dark bands, moving to smaller wavevectors (angles) for larger wavelengths. The measured BFP spectra (Fig.~\ref{fig:figure4}F and G) revealed  the trend of the stopband with varied angles and wavelengths. Further,  the differently size particles displayed different stopband positions.

The wavevectors imaged in the BFP correspond to light incident at different angles on the plane parallel to the substrate, which corresponds to the (111) plane for the assembled FCC crystal structure. The interference of the scattered light from the set of parallel (111) crystallographic planes give rise to Bragg peaks, analogous to the Bragg peaks seen in X-ray diffraction experiments on atomic crystal lattices. The Bragg peaks correspond to high reflectivity, and result in lower intensity of collected light. The Bragg condition for such an optical system is given as \cite{gonzalez2012linear} :
\begin{equation}\label{eq2}
 2 \, d_\text{hkl} \, n_\text{0} \, cos(\theta) = m \lambda
\end{equation}


where $d_\text{hkl}$ is the interplanar spacing, $n_\text{0}$ is the effective refractive index, $\theta$ is the angle of emission, and $\lambda$ is the wavelength. We use $d_\text{hkl} = d_\text{111} = 0.816 \, D$, which corresponds to the spacing of the (111) planes of the FCC crystal, with $D$ being the diameter of the particles used. The effective refractive index $n_\text{0}$ was approximated as $n_\text{0} = f_\text{p}n_\text{p} + f_\text{m}n_\text{m}$ \cite{aspnes1982optical}, where $f_\text{p}$ and $f_\text{m}$ are the filling fractions of the particles and medium, respectively, and $n_\text{p}$ and $n_\text{m}$ are their corresponding refractive indices.
The stopbands obtained in our experiments were found to be in good agreement with the predictions of the Bragg equation (plotted in Fig.~\ref{fig:figure4} as dotted lines). Further, the trend of the wavelength dependence of the stopband obtained from the BFP spectra also matched well with the Bragg predictions. The presence of the stop band is also an indicator of crystallinity, that is otherwise hard to resolve for the smaller $390\,\si{\nano\metre}$ and $400\,\si{\nano\metre}$ PS particles.

Since the photonic stopband emerges from the crystallinity of the structure probed by the flourescence at the focus of the laser, it disappers when the crystal structure is destroyed at the center by high temperatures. When the laser power was cycled between a high and low heating power ($P_\text{0}$ and $P_\text{1}$, as described in the earlier section) with a pulse duration of 4s, the oscillating crystallinity at the center resulted in an oscillating photonic stopband Fig.~\ref{fig:figure4}. The photonic stopbands disappeared whenever the crystal was destroyed in the center by the formation of the dynamic toroidal structure, but reappeared whenever the colloids were collapsed back into a crystal structure. The presence of the stopband in the collapsed structures elucidates that the crystallinity is still preserved in structures formed by the rapid quenching process.
\begin{figure*}[!htb]
    \centering
    \includegraphics[width=0.9 \textwidth]{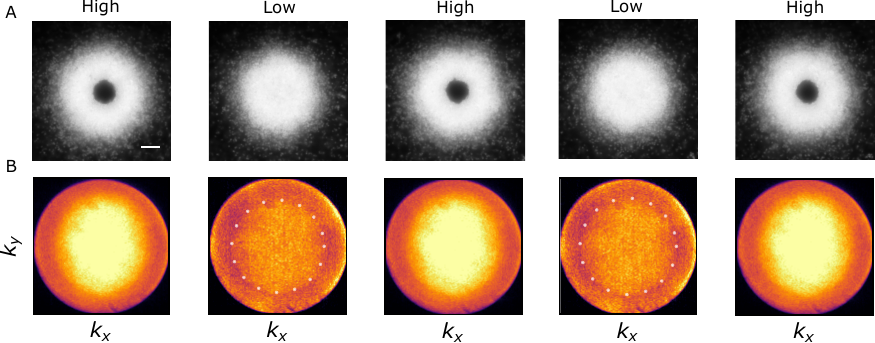}
    \caption{\textbf{Modulation of photonic stopband} (A) Microscopy images of the structures at the peak of the high and low intensity pulses.  (B) Photonic stopband at the peak of the high and low intensity pulses. The period of the pulse used was $8\, \si{\second} $. $345 \, \si{\nano\metre}$ PS particles and $660 \si{\nano\metre}$ band-pass filter was used.}\label{fig:figure4}
\end{figure*}
Inspite of the rapid nature of the quenching process, the diffusion of the nanoparticles during the time-scale of the collapse is still high enough to allow for structural relaxation required to prevent the freezing of defects typically associated with quenching processes. We found that employing these accelerated growth dynamics (Fig.~\ref{fig:figure2}B and C) does not adversely impact the crystallinity of the final structures, enhancing speed without compromising quality.
\\

\noindent
\textbf{Discussion}
When the tightly focussed laser spot (beam waist $250\,\si{\nano\metre}$) is absorbed  by the gold layer, it generates heat locally. This results in a stationary inhomogeneous temperature profile in water within a time duration of a few tens of microseconds \cite{simon2023thermoplasmonic} due to the high thermal diffusivity of water ($\alpha_\text{water} = 0.14 \cdot 10^{-6} \, \si{\metre\squared\per\second} $). The resulting temperature gradients can drive a flow of particles towards the heated area due to various thermofluidic effects.

The temperature gradients at the substrate-solvent interface perturb the liquid-solid interactions and result, among other effects, in thermoosmotic flows \cite{franzl2022hydrodynamic}. The flow field inside the liquid film can then be calculated from the hydrodynamic boundary condition at the gold liquid interface with the slip velocity
    \begin{equation}\label{eq3}
    \mathbf{u}_\text{TO, slip} = - \frac{1}{\eta} \int_{0}^{\infty} {zh(z)} \, \text{dz} \frac{\nabla T}{T} = \chi_\text{AuL}  \frac{\nabla T}{T}
    \end{equation}
where $\eta$ is the viscosity of the liquid, $h(z)$ is the excess free enthalpy density, $T$ is the temperature, $\nabla T$ is the temperature gradient at the interface. $\chi_\text{AuL}$ is the thermo-osmotic coefficient at the Gold-liquid interface, which takes a value of $10 \times 10^{-10}\, \si{\metre\squared\per\second}$ as previously measured for the gold water interface \cite{franzl2022hydrodynamic}.

The temperature gradients across the solute particles and molecules in the liquid also cause a thermophoretic drift current density \cite{piazza2008thermophoresis, niether2019thermophoresis}. The corresponding drift velocity $\mathbf{v}_\text{TP}$ of the objects undergoing thermophoresis is described by the equation:
    \begin{equation}\label{eq4}
    \mathbf{v}_\text{TP} = - \frac{2}{3} \chi_\text{SL}  \frac{\nabla T}{T}  = - D_\text{T,i} \nabla T
    \end{equation}
where $\chi_\text{SL}$ is the osmotic coefficient at the solid-liquid interface, which gives rise to the thermophoretic mobility of the species $D_\text{T,i}$, where $i=(C=\text{PS},P=\text{PEG})$, which take on values $D_\text{T,C} = 0.3 \, \si{\micro\metre\squared\per\kelvin\per\second}$ \cite{franzl2022hydrodynamic} and $D_\text{T,P} = 0.84 \, \si{\micro\metre\squared\per\kelvin\per\second}$ \cite{chan2003soret}. The thermophoresis alters the distribution of the PEG molecules, which reaches a steady state when the thermophoretic and diffusive molecule fluxes balance, resulting in a concentration profile described by eq. \ref{eq5}.
\begin{equation}\label{eq5}
c_\text{P}(\mathbf{r}) = c_\text{0,P} \, \exp(-S_\text{T,P} \Delta T(\mathbf{r}))
\end{equation}

where $c_\text{0,P}$ is the initial concentration of PEG, $S_\text{T,P}=D_\text{T,P}/D_\text{P}=0.05\,\si{\kelvin^{-1}}$ is the Soret coefficient of the PEG molecules \cite{chan2003soret}. The thermally generated concentration profile results in an excess osmotic pressure that acts on the PS particles. For an interaction potential $\phi$ between the PS particles and solute molecules the osmotic pressure can be calculated by
    \begin{equation}\label{eq6}
    \Pi(\mathbf{r}) = c_\text{P}(\mathbf{r}) k_{B} T \left (e^{-\frac{\phi}{k_{B}T}} - 1\right)
    \end{equation}
where $k_\text{B}$ is the Boltzmann constant, and yields a drift of the particles towards the heated spot \cite{maeda2011thermal,maeda2012effects,jiang2009manipulation}. The drift velocity of the particles due to this depletion interaction is then given by \cite{wurger_thermal_2010}:
    \begin{equation}\label{eq7}
    \mathbf{v}_\text{D}(\mathbf{r}) = \frac{k_\text{B}}{3 \eta} R_\text{P}^{2} c_\text{P}(\mathbf{r})  (T S_\text{T,P} - 1) \nabla T
    \end{equation}
where $R_\text{P}=3.62\,\si{\nano\metre}$ is the radius of gyration of the PEG molecules used.
\begin{figure*}[!htb]
    \centering
    \includegraphics[width=0.9 \textwidth]{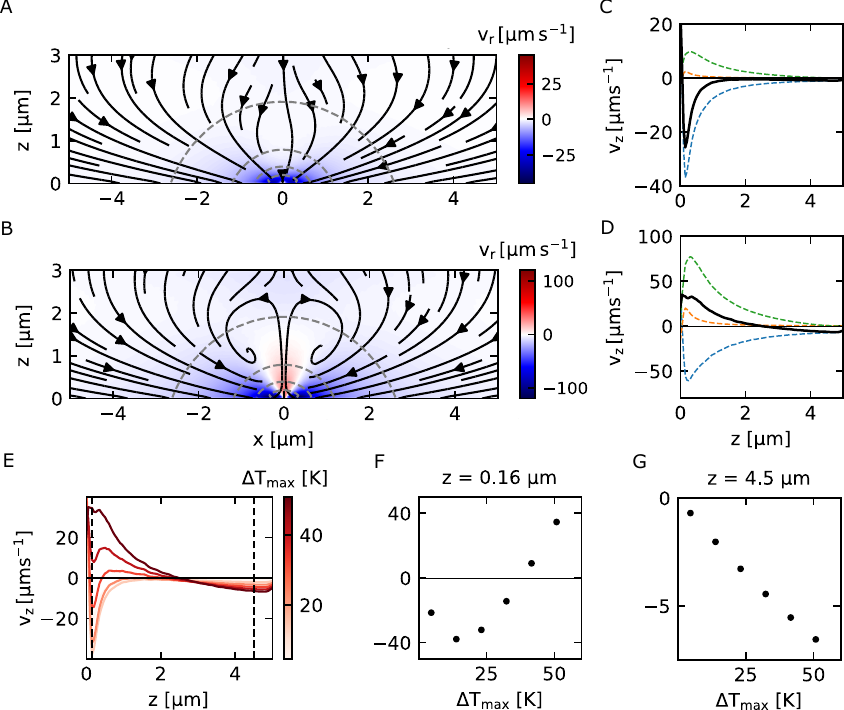}
    \caption{\textbf{Thermofluidic velocity fields} Velocity fields were computed, corresponding to laser power (A) $P_0 = 0.26 \,\si{\milli\watt}$ and (B) $P_1 = 2.16 \,\si{\milli\watt}$. The colour map represents the radial velocities. A repulsive region opens up in the center for the higher laser power. (C and D) shows the vertical component of the velocities at $x=0$ for the corresponding powers (blue = Depletion; red = thermophoresis; green = depletion; black = total). (E) Total velocities for different temperature increments. (F and G) show the velocities for the line profile at z=0.16 ${\mu m}$ and z=4.5 ${\mu m}$ respectively.}\label{fig:figure5}
\end{figure*}
The assembly of the colloidal particles as observed in the experiment depends on the interplay of these thermofluidic effects. According to Eqn. \ref{eq3}, \ref{eq4} and \ref{eq7} all of these drifts depend on the gradient of the temperature field, yet, eq. \ref{eq3} only refers to the temperature gradient along the solid-liquid boundary. The thermophoretic and the diffusiophoretic drifts (Eqn. \ref{eq4} and \ref{eq7}) follow the symmetry of the temperature field. The velocities are perpendicular to the temperature isotherms and can be either along the temperature gradient (attractive to the heat source) in the case of diffusiophoresis induced drift (Eq. \ref{eq7}) or against the temperature gradient (repulsive with positive $D_\text{T}$) in the case of thermophoresis (Eq. \ref{eq4}). Thus any change in the heating power and the maximum temperature would only yield a quantitative change in the distribution of molecules and particles in the system, but no qualitative change as observed in the experiment when changing from low to high powers.

The thermoosmotic boundary flow described by Eq. \ref{eq3}, however, resembles a boundary slip flow that drives the flow field in the liquid film. The drag forces acting on the particles and molecules therefore do not follow the temperature gradient direction everywhere in the liquid. In particular, the flow field shows an upward flow in the z-direction above the hot spot and a radial outflow in the centre of the liquid film due to mass conservation. The thermoosmotic flow therefore breaks the symmetry of the temperature field and will give rise to new qualitative behaviour if the flow generates drift velocities on the particles that are of the same order of magnitude or higher than the other drift velocities.

This qualitative change can be substantiated by a numerical calculation of the total thermofluidic drift velocities (Eq. \ref{eq3}, \ref{eq4}, and \ref{eq7}) two of the different experimental optical heating powers. Fig.~\ref{fig:figure5}A and B show the calculated drift field for the two used powers $P_0$ and $P_1$. While for $P_0$ all drift velocities point radially inwards to the heat source, Fig.\ref{fig:figure5}B reveals a reversal of the sign of the radial drift speed $v_r$ for $P_1$.

Fig.~\ref{fig:figure5}C and D decomposes the contributions showing the normal component $v_z(z)$ of the drift velocities for a point-like particle located at $x=0$. When $\Delta T_\text{max} = 6\, \si{\kelvin}$  for $P_0$, the total velocity vectors are always directed towards the heated spot. As visible in Fig.\ref{fig:figure5}C the attraction to the hot spot is caused by the PEG gradient while thermo-osmotic flow and thermophoresis play only a minor role.

Switching to $P_1$, i.e to $\Delta T_\text{max} = 48.8\, \si{\kelvin}$ causes a stronger contribution of the PEG gradient to $v_z$ but also stronger thermo-osmotic flows. The total speed $v_z$ therefore points away from the heated spot at the center (seen as positive velocities) up to a distance of $z=2.2\, \si{\micro\metre}$, beyond which it continues pointing in the direction of the heated spot. Yet, the zero drift speed in the vertical direction is not a stable point in the flow field due to the thermo-osmotic outflow in the middle of the film. Due to this outflow, a toroid-like structure appears in which the particles are confined and which is clearly visible in the experiment 
(see \ref{SI} movie)

Fig.~\ref{fig:figure5}E highlights this transition from an attractive to a repulsive flow in the vertical z-direction again for different maximum temperature increments $\Delta T_{\rm max}$. Plotting $v_z$ at $z=0.16\,\si{\micro\metre}$ (Fig.~\ref{fig:figure5}F) indicates that initially the attraction is enhanced and then reverses to a repulsive behavior ($v_z>0\,\si{\micro\metre\per\second}$) at $\Delta T_{\rm max}=32.5\,\si{\kelvin}$ for this position. Yet, at $z=4.5\, \si{\micro\metre}$ we still find a linear increase of the attractive speed $v_{z}<0\,\si{\micro\metre\per\second}$ towards the heated region from the numerical calculation (Fig. \ref{fig:figure5}G). The heated region therefore continues to attract colloidal particles, but the colloids do not reach the central hot spot, instead becoming trapped in the toroid structure as found in the experiment as well.

The flow speeds obtained from the numerical calculations of the different contributions can now be used to quantify the crystal growth. For this purpose we consider the growth at low power where all particles drifts are towards the heart source (Fig.~\ref{fig:figure5}A).  For the measured area above the interface, the inflow of particles due to the drift tangential to the interface is important. The phenomenological equation for this particle inflow can be written as:
\begin{equation}\label{eq9}
\mathbf{j}_\text{c,t} = \mathbf{v}_\text{c,t} c_\text{c} 
\end{equation}

where $\mathbf{v}_\text{c,t} $ corresponds to the tangential drift velocities of the colloids that can be found by averaging the drift velocities from the underlying thermofluidic processes (eq. \ref{eq3}, \ref{eq4}, \ref{eq7}) obtained from numerical calculations over the distance in $\text{z}$ explored by the colloids between successive camera frames $\Delta z = 2R_\text{c}+D_\text{z} = 0.45 \, \si{\micro\metre}$, where $R_\text{c} = 0.17 \, \si{\micro\metre}$ is the radius of the colloidal particle and $D_\text{z} = 0.1 \, \si{\micro\metre}$ is the diffusion length in $z$ of the colloids at the interface in this time interval.

Based on the constant influx of particles to the central region at low heating power, the conservation of particle number dictates the growth of the observed crystal by a power law (eq. \ref{eq1}) with an exponent of $\alpha=2/3$. The rate constant quantifying the crystal growth is related to the different drift velocities as:

\begin{equation}\label{eq10}
k = \frac{4\, \pi^{5/2} \, R_\text{c}^3  \, v_\text{c,t}^\text{0} \, c_\text{c}}{\eta_\text{pack}}
\end{equation}

where $v_\text{c,t}^\text{0}$ is the magnitude of the tangential velocity at the edge of the heat source (taken as $x=0.3 \, \si{\micro\metre}$), $c_\text{c} = 0.8 \, \si{\per\micro\metre\cubed}$ is the number density, and $\eta=0.74$ is the packing fraction of the colloidal particles. The derived power law is similar to the Ostwald coarsening process \cite{bray1994theory} and was also observed in experiments where passive particles were assembled by phoretic fluxes induced by active particles \cite{boniface2024clustering}. For the lower temperature increments $\Delta T_\text{max} = 6\, \si{\kelvin}$, we calculate $v_\text{c,t}^\text{max} = 25.0 \, \si{\micro\meter\per\second}$, resulting in a growth constant $k = 9.6 \, \si{\micro\meter\cubed\per\second}$, in close agreement with the value fitted from the experimental curve (see Fig.~\ref{fig:figure2}A).

For a temperature increment of $\Delta T_\text{max} = 48.8 \, \si{\kelvin}$, which is 8.13 times higher, the growth constant $k$ would increase proportionally since velocities outside the central region scale linearly (see Fig.~\ref{fig:figure5}G). This would yield an area ${8.13}^{2/3} = 4.04$ times larger than the previous case at the same time point. In our experiments, the measured growth rate $k$, calculated from the final area, was 3.25 times higher - slightly lower than the model's prediction. 
Despite this small difference, our model successfully captures how stronger hydrodynamic fluxes affect structure size at higher temperature increments and provides a framework for quantifying particle accumulation. The model unifies various thermofluidic processes through the velocity parameter $v_\text{c,t}^\text{0}$, which emerges as the key factor controlling growth in a given sample and is experimentally accessible via the laser power.


With the quantitative understanding of the interplay of temperature-related forces and flows at a local level - including thermophoresis, thermo-osmosis, and temperature-based depletion - we can guide individual components to come together and form structures. This allows us to accurately control the assembly dynamics and kinetics by varying the input energy or the assembly components. Changing the input energy via the laser power not only changes the kinetics of the assembly but also the nature of the structures, with new dynamic structures, unique to non-equilibrium assembly, emerging when we break their symmetry via the temperature.

The ability to precisely control local assembly through temperature gradients opens new possibilities for creating adaptive materials with programmable properties. The interplay between different thermofluidic effects demonstrated here suggests that even more complex assembly modes could emerge when working with mixtures of particles of different sizes, shapes, or surface properties. For instance, particles with different thermophoretic responses could self-sort into distinct regions, while asymmetric particles might orient themselves along the temperature gradients, potentially leading to hierarchical structures. The dynamic nature of the heating could be exploited to create temporal patterns in the assembly process, enabling sequential organization of different components or the formation of gradient structures.

While the dynamic structures reported here arise due to thermofluidic interactions of simple spherical colloids, many more unique dynamic structures could arise from additional interactions in more complex colloids or mixtures. For example, particles with different surface chemistries might exhibit competing attractions and repulsions, leading to the formation of novel spatiotemporal patterns.

Furthermore, this assembly method could be extended beyond colloidal systems. New modes of accumulation could be implemented for molecules or analytes of interest, potentially enabling applications in chemical sensing or molecular separation. In biological contexts, components such as cells that are inherently active could organize into even more complex and dynamic patterns, suggesting possible applications in tissue engineering or research into cellular organization. Precise spatial and temporal control over temperature gradients could be particularly valuable for coordinating the assembly of temperature-sensitive biological components or creating biomimetic structures that respond to environmental cues.

\section*{Conclusion}
We have demonstrated the assembly of 3D structures maintained by continuous energy dissipation through localized optical heating. By quantifying the interplay of different temperature-induced processes - thermophoresis, thermo-osmosis and temperature-based depletion - we were able to explain both the growth dynamics of crystalline structures and the emergence of novel out-of-equilibrium toroidal assemblies. The assembled structures exhibited reconfigurable photonic properties, manifested in tunable stopbands that could be modulated by precise control of the temperature field. Remarkably, these 3D photonic crystals could be assembled within minutes, much faster than conventional approaches \cite{deubel2004direct,von2013bottom,cong2003colloidal}. The rapid assembly combined with the ability to dynamically reconfigure the structures through temperature control demonstrates the potential of thermofluidic assembly for the creation of adaptive functional materials. This approach opens up new possibilities for the design of dynamic structures that can respond to external stimuli - a capability that cannot be readily achieved with conventional top-down or equilibrium self-assembly techniques.

\section*{Materials and Methods} \label{Methods}
\textbf{Sample preparation:} The sample consists of a liquid film of $5 \,\si{\micro\metre}$ thickness containing a solution of polystyrene (PS) nanoparticles in 7 $\%$ (w/v) PEG-6000 ($11.67$ mM) (along with $5\, \si{\micro\metre}$ spacers). A mixture of Rhodamine-6G ($10^{-6}$ M) and Nile Blue ($10^{-5}$ M) was further added to the samples where a fluorescent signal was neccessary for the measurement of photonic stopbands. The liquid sample lies between two glass coverslips, with the bottom coverslip coated with a 50\, \si{\nano\metre} thin Gold film to facilitate laser induced heating. Polystyrene particles of 3 different sizes were purchased from Microparticles GmbH and Dynamic Light Scattering (Malvern Zetasizer Ultra) measurements of the particles revealed average diameters of $345 \, \si{\nano\metre}$, $400 \, \si{\nano\metre}$ and $442 \, \si{\nano\metre}$ with polydispersities of 0.006, 0.040, and 0.123 respectively. The height of the liquid film maintained by the spacers was several times the diameter of the nanoparticles, providing space for growth in 3 dimensions. Further, the dense colloidal solution used ($ 0.2 \, \% $ (w/v)) facilitated the assembly of a larger and 3-dimensional structure.
%
%
\textbf{Experimental setup:} The experimental setup consists of an inverted microscope (Olympus, IX71) with an oil-immersion objective lens (Olympus, UPlanApo x100, NA $0.5-1.35$) focusing a CW 532 nm laser beam (CNI, MGL-III-532) onto the sample. The sample is kept on a high precision piezoelectric stage (Physik instrumente, P-733.3 piezo nanopositioner) that allows fine movement in all 3 dimensions. The laser beam passes through an Acousto-Optic Deflector (AA Opto-Electronic, DTSXY-400-532), which allows for the accurate positioning and power modulation of the focused laser spot. A LED white light source (Thorlabs, SOLIS-3C) and a Dark-field condenser (Olympus, U-DCW, NA 1.2-1.4) provide dark-field illumination to the sample. After the back-reflected laser-wavelength is rejected from the collected signal using a notch-filter, the light is projected onto the camera (EMCCD, Andor iXon), which images the real plane. The excited fluorescence of the dye mixture is probed in the Fourier plane. The Fourier plane (FP) imaging and spectrosopy is performed by a 4f relay configuration \cite{kurvits2015comparative}, which images the Back focal plane of the Objective. The FP image is projected onto an adjustable slit. When the slit is fully open, it allows the camera (sCMOS) to capture the image of the Fourier plane. The FP image contains angle resolved information of the fluorescence signal that passes through the sample. Wavelength dependent FP imaging is perfored using band-pass filters centered around 565 nm (565WB20), 590 nm (590BP35) and 660 nm (660BP20) wavelength. To obtain additional spectroscopic information, the light is projected onto the camera through a blazed grating (which is flipped on), while using a narrow slit width.

\section*{Acknowledgements}
The authors acknowledge financial support by the Federal Ministry for Economic Affairs and Energy based on a resolution of the German Bundestag (BMWi, STARK programme, project number 46SKD023X to F.C.), cofinanced from tax revenues on the basis of the budget passed by the Saxon state parliament (SMWK). We thank A. Kramer for proof-reading the manuscript.

\section*{Supplementary Information} \label{SI}
The Supporting Information is available free of charge.


\begin{thebibliography}{10}

\bibitem{alberts2017molecular}
Bruce Alberts.
\newblock {\em Molecular biology of the cell}.
\newblock Garland Science, 2017.

\bibitem{hess2017non}
H~Hess and Jennifer~L Ross.
\newblock Non-equilibrium assembly of microtubules: from molecules to
  autonomous chemical robots.
\newblock {\em Chemical Society Reviews}, 46(18):5570--5587, 2017.

\bibitem{huh2020exploiting}
Ji-Hyeok Huh, Kwangjin Kim, Eunji Im, Jaewon Lee, YongDeok Cho, and Seungwoo
  Lee.
\newblock Exploiting colloidal metamaterials for achieving unnatural optical
  refractions.
\newblock {\em Advanced Materials}, 32(51):2001806, 2020.

\bibitem{rocha2020role}
Brunno~C Rocha, Sanjib Paul, and Harish Vashisth.
\newblock Role of entropy in colloidal self-assembly.
\newblock {\em Entropy}, 22(8):877, 2020.

\bibitem{sciortino2019entropy}
Francesco Sciortino.
\newblock Entropy in self-assembly.
\newblock {\em La Rivista del Nuovo Cimento}, 42:511--548, 2019.

\bibitem{li2021colloidal}
Zhiwei Li, Qingsong Fan, and Yadong Yin.
\newblock Colloidal self-assembly approaches to smart nanostructured materials.
\newblock {\em Chemical Reviews}, 122(5):4976--5067, 2021.

\bibitem{boles2016self}
Michael~A Boles, Michael Engel, and Dmitri~V Talapin.
\newblock Self-assembly of colloidal nanocrystals: from intricate structures to
  functional materials.
\newblock {\em Chemical Reviews}, 116(18):11220--11289, 2016.

\bibitem{kim2011self}
Shin-Hyun Kim, Su~Yeon Lee, Seung-Man Yang, and Gi-Ra Yi.
\newblock Self-assembled colloidal structures for photonics.
\newblock {\em NPG Asia Materials}, 3(1):25--33, 2011.

\bibitem{zhu2017assembly}
Jian Zhu and Mark~C Hersam.
\newblock Assembly and electronic applications of colloidal nanomaterials.
\newblock {\em Advanced Materials}, 29(4):1603895, 2017.

\bibitem{wintzheimer2018supraparticles}
Susanne Wintzheimer, Tim Granath, Maximilian Oppmann, Thomas Kister, Thibaut
  Thai, Tobias Kraus, Nicolas Vogel, and Karl Mandel.
\newblock Supraparticles: functionality from uniform structural motifs.
\newblock {\em ACS Nano}, 12(6):5093--5120, 2018.

\bibitem{zhang2009colloidal}
Gang Zhang and Dayang Wang.
\newblock Colloidal lithography—the art of nanochemical patterning.
\newblock {\em Chemistry--An Asian Journal}, 4(2):236--245, 2009.

\bibitem{prigogine2017non}
Ilya Prigogine.
\newblock {\em Non-equilibrium statistical mechanics}.
\newblock Courier Dover Publications, 2017.

\bibitem{heinen2015celebrating}
Laura Heinen and Andreas Walther.
\newblock Celebrating soft matter's 10th anniversary: Approaches to program the
  time domain of self-assemblies.
\newblock {\em Soft Matter}, 11(40):7857--7866, 2015.

\bibitem{arango2019self}
A~Arango-Restrepo, D~Barrag{\'a}n, and JM~Rubi.
\newblock Self-assembling outside equilibrium: emergence of structures mediated
  by dissipation.
\newblock {\em Physical Chemistry Chemical Physics}, 21(32):17475--17493, 2019.

\bibitem{Trivedi.2022}
Manish Trivedi, Dhruv Saxena, Wai~Kit Ng, Riccardo Sapienza, and Giorgio Volpe.
\newblock {Self-organized lasers from reconfigurable colloidal assemblies}.
\newblock {\em Nature Physics}, 18(8):939--944, 2022.

\bibitem{van2020out}
Bas~GP van Ravensteijn, Ilja~K Voets, Willem~K Kegel, and Rienk Eelkema.
\newblock Out-of-equilibrium colloidal assembly driven by chemical reaction
  networks.
\newblock {\em Langmuir}, 36(36):10639--10656, 2020.

\bibitem{van2017dissipative}
Susan~AP van Rossum, Marta Tena-Solsona, Jan~H van Esch, Rienk Eelkema, and Job
  Boekhoven.
\newblock Dissipative out-of-equilibrium assembly of man-made supramolecular
  materials.
\newblock {\em Chemical Society Reviews}, 46(18):5519--5535, 2017.

\bibitem{grzelczak2010directed}
Marek Grzelczak, Jan Vermant, Eric~M Furst, and Luis~M Liz-Marz{\'a}n.
\newblock Directed self-assembly of nanoparticles.
\newblock {\em ACS Nano}, 4(7):3591--3605, 2010.

\bibitem{bharti2015assembly}
Bhuvnesh Bharti and Orlin~D Velev.
\newblock Assembly of reconfigurable colloidal structures by multidirectional
  field-induced interactions.
\newblock {\em Langmuir}, 31(29):7897--7908, 2015.

\bibitem{ma2013formation}
Fuduo Ma, David~T Wu, and Ning Wu.
\newblock Formation of colloidal molecules induced by alternating-current
  electric fields.
\newblock {\em Journal of the American Chemical Society}, 135(21):7839--7842,
  2013.

\bibitem{al2020magnetic}
A~Al~Harraq, JG~Lee, and B~Bharti.
\newblock Magnetic field--driven assembly and reconfiguration of multicomponent
  supraparticles.
\newblock {\em Science Advances}, 6(19):eaba5337, 2020.

\bibitem{bregulla2016thermo}
Andreas~P Bregulla, Alois W{\"u}rger, Katrin G{\"u}nther, Michael Mertig, and
  Frank Cichos.
\newblock Thermo-osmotic flow in thin films.
\newblock {\em Physical Review Letters}, 116(18):188303, 2016.

\bibitem{franzl2022hydrodynamic}
Martin Fr{\"a}nzl and Frank Cichos.
\newblock Hydrodynamic manipulation of nano-objects by optically induced
  thermo-osmotic flows.
\newblock {\em Nature Communications}, 13(1):656, 2022.

\bibitem{jiang2009manipulation}
Hong-Ren Jiang, Hirofumi Wada, Natsuhiko Yoshinaga, and Masaki Sano.
\newblock Manipulation of colloids by a nonequilibrium depletion force in a
  temperature gradient.
\newblock {\em Physical Review Letters}, 102(20):208301, 2009.

\bibitem{maeda2011thermal}
Yusuke~T Maeda, Axel Buguin, and Albert Libchaber.
\newblock Thermal separation: interplay between the soret effect and entropic
  force gradient.
\newblock {\em Physical Review Letters}, 107(3):038301, 2011.

\bibitem{liu2021opto}
Shaofeng Liu, Linhan Lin, and Hong-Bo Sun.
\newblock Opto-thermophoretic manipulation.
\newblock {\em ACS Nano}, 15(4):5925--5943, 2021.

\bibitem{kurvits2015comparative}
Jonathan~A Kurvits, Mingming Jiang, and Rashid Zia.
\newblock Comparative analysis of imaging configurations and objectives for
  fourier microscopy.
\newblock {\em JOSA A}, 32(11):2082--2092, 2015.

\bibitem{wagner2013fast}
Rebecca Wagner and Frank Cichos.
\newblock Fast measurement of photonic stop bands by back focal plane imaging.
\newblock {\em Physical Review B}, 87(16):165438, 2013.

\bibitem{gonzalez2012linear}
Luis Gonz{\'a}lez-Urbina, Kasper Baert, Branko Kolaric, Javier
  P{\'e}rez-Moreno, and Koen Clays.
\newblock Linear and nonlinear optical properties of colloidal photonic
  crystals.
\newblock {\em Chemical reviews}, 112(4):2268--2285, 2012.

\bibitem{aspnes1982optical}
David~E Aspnes.
\newblock Optical properties of thin films.
\newblock {\em Thin solid films}, 89(3):249--262, 1982.

\bibitem{simon2023thermoplasmonic}
David~J Simon, Felix Hartmann, Tobias Thalheim, and Frank Cichos.
\newblock Thermoplasmonic manipulation for the study of single polymers and
  protein aggregates.
\newblock {\em Macromolecular Chemistry and Physics}, 224(14):2300060, 2023.

\bibitem{piazza2008thermophoresis}
Roberto Piazza and Alberto Parola.
\newblock Thermophoresis in colloidal suspensions.
\newblock {\em Journal of Physics: Condensed Matter}, 20(15):153102, 2008.

\bibitem{niether2019thermophoresis}
Doreen Niether and Simone Wiegand.
\newblock Thermophoresis of biological and biocompatible compounds in aqueous
  solution.
\newblock {\em Journal of Physics: Condensed Matter}, 31(50):503003, 2019.

\bibitem{chan2003soret}
Joey Chan, Jesse~J Popov, Stacey Kolisnek-Kehl, and Derek~G Leaist.
\newblock Soret coefficients for aqueous polyethylene glycol solutions and some
  tests of the segmental model of polymer thermal diffusion.
\newblock {\em Journal of Solution Chemistry}, 32:197--214, 2003.

\bibitem{maeda2012effects}
Yusuke~T Maeda, Tsvi Tlusty, and Albert Libchaber.
\newblock Effects of long dna folding and small rna stem--loop in
  thermophoresis.
\newblock {\em Proceedings of the National Academy of Sciences},
  109(44):17972--17977, 2012.

\bibitem{wurger_thermal_2010}
Alois Würger.
\newblock Thermal non-equilibrium transport in colloids.
\newblock {\em Reports on Progress in Physics}, 73(12):126601, November 2010.
\newblock Publisher: IOP Publishing.

\bibitem{bray1994theory}
Alan~J Bray.
\newblock Theory of phase-ordering kinetics.
\newblock {\em Advances in Physics}, 43(3):357--459, 1994.

\bibitem{boniface2024clustering}
Dolachai Boniface, Sergi~G Leyva, Ignacio Pagonabarraga, and Pietro Tierno.
\newblock Clustering induces switching between phoretic and osmotic propulsion
  in active colloidal rafts.
\newblock {\em Nature Communications}, 15(1):5666, 2024.

\bibitem{deubel2004direct}
Markus Deubel, Georg Von~Freymann, Martin Wegener, Suresh Pereira, Kurt Busch,
  and Costas~M Soukoulis.
\newblock Direct laser writing of three-dimensional photonic-crystal templates
  for telecommunications.
\newblock {\em Nature Materials}, 3(7):444--447, 2004.

\bibitem{von2013bottom}
Georg von Freymann, Vladimir Kitaev, Bettina~V Lotsch, and Geoffrey~A Ozin.
\newblock Bottom-up assembly of photonic crystals.
\newblock {\em Chemical Society Reviews}, 42(7):2528--2554, 2013.

\bibitem{cong2003colloidal}
Hailin Cong and Weixiao Cao.
\newblock Colloidal crystallization induced by capillary force.
\newblock {\em Langmuir}, 19(20):8177--8181, 2003.

\end{thebibliography}
%

\end{document}